\documentclass[aps,amsmath,twocolumn,amssymb,floatfix,showpacs,superscriptaddress,nofootinbib,longbibliography]{revtex4-1}
\usepackage{braket}
\usepackage[dvipsnames]{xcolor}
\usepackage{float}
\usepackage{subfigure}
\usepackage{tikz}
\usepackage[colorlinks=true,linktoc=page,linkcolor=OliveGreen,citecolor=purple,urlcolor=violet]{hyperref}

\mathchardef\mhyphen="2D 

\newcommand{\ie}{{i.e.,\,\,}}
\newcommand{\eg}{{e.g.,~}}

\newcommand\bea{\begin{eqnarray}}
	\newcommand\eea{\end{eqnarray}}
\newcommand\beq{\begin{equation}}  
	\newcommand\eeq{\end{equation}}

\usepackage[normalem]{ulem}
\definecolor{lime}{HTML}{A6CE39}
\usepackage{sidecap,tikz}
\DeclareRobustCommand{\orcidicon}{\hspace{-1.0mm}
	\begin{tikzpicture}
		\draw[lime, fill=lime] (0.0,0.0) 
		circle [radius=0.15] 
		node[white] {{\fontfamily{qag}\selectfont \tiny \,ID}};
		\draw[white, fill=white] (-0.0525,0.095) 
		circle [radius=0.007];
	\end{tikzpicture}
	\hspace{-3.0mm}
}
\foreach \x in {A, ..., Z}{\expandafter\xdef\csname orcid\x\endcsname{\noexpand\href{https://orcid.org/\csname orcidauthor\x\endcsname}{\noexpand\orcidicon}}
}

\AtBeginDocument{%
	\newwrite\bibnotes
	\def\bibnotesext{Notes.bib}
	\immediate\openout\bibnotes=\jobname\bibnotesext
	\immediate\write\bibnotes{@CONTROL{REVTEX41Control}}
	\immediate\write\bibnotes{@CONTROL{%
			apsrev41Control,author="08",editor="1",pages="1",title="1",year="1"}}
	\if@filesw
	\immediate\write\@auxout{\string\citation{apsrev41Control}}%
	\fi
}%

\begin{document}

\title{Hierarchy of topological superconductivity generated via heterostructures of unconventional $p$-wave magnets}

\author{Koushik R. Das \orcidA{}}
\email{koushik.das@iopb.res.in}
\affiliation{Institute of Physics, Sachivalaya Marg, Bhubaneswar-751005, India}

\author{Arijit Saha \orcidB{}}
\email{arijit@iopb.res.in}
\affiliation{Institute of Physics, Sachivalaya Marg, Bhubaneswar-751005, India}
\affiliation{Homi Bhabha National Institute, Training School Complex, Anushakti Nagar, Mumbai 400094, India}

\begin{abstract}
	A theoretical framework is proposed to engineer both first and second-order topological superconducting phases in a two-dimensional (2D) heterostructure, consisting of a quantum spin Hall insulator (QSHI) and an unconventional $p$-wave magnet in presence of proximity-induced $s$-wave superconducting pairing. Our analysis establishes that the transitions between the trivial and topological superconducting (TSC) phases can be regulated though the parameters of $p$-wave magnet. Presence of chiral symmetry leads to the characterization of both types of TSC phases by the respective invariants, one-dimensional winding number and quadrupolar winding number. These results are supplemented by an analytical effective low-energy edge theory that yields a deeper insight into the emergence of the different topological phases of the system. Bulk pairing analysis reveals the competition between the effective $(p_x+p_y)$ and $(p_x+ip_y)$ type pairings that are governed by the intrinsic spin-orbit coupling inherited to the QSHI and spin-split bands of the $p$-wave magnet, respectively. 
\end{abstract}

\maketitle

{\it{\textcolor{blue}{Introduction}}}:-
	The quest to engineer Majorana Fermions in solid state systems has been a central focus of modern condensed matter physics. Since the theoretical prediction of Majorana zero modes (MZMs) in a one-dimensional (1D) spinless $p$-wave superconductor by Kitaev in 2001~\cite{kitaev2001unpaired}, Majorana physics has emerged as a major field of research~\cite{ivanov2001non,kitaev2009periodic,elliott2015colloquium,lutchyn2010majorana,oreg2010helical,lutchyn2018majorana,yan2018majorana,alicea2012new,stanescu2013majorana} for the last two decades. MZMs are neutral-charge zero-energy quasiparticle excitations in superconducting systems 
	that respect non-Abelian braiding statistics~\cite{beenakker2013search,amorim2015majorana,clarke2017probability,malciu2018braiding,sanno2021ab,mascot2023many,zhang2019majorana,stenger2021simulating,ma2026direct,buss2026braiding,stephan2024dis,boross2024dot,mazur2025maj}, and therefore carry the potential to revolutionize fault-tolerant topological quantum computing~\cite{nayak2008non,marra2022majorana,aasen2016milestones,sarma2015majorana,harle2023observing}. Consequently, the emergence of such MZMs in topological superconductors has attracted significant attention in the scientific community 
	in recent times~\cite{flensberg2021engineered,sato2017topological,li2019exploring,val2022topological}. 
	Since the early proposals of Majorana bound states engineered in topological insulator surface states in presence of proximity-induced superconductivity and ferromagnetic insulator~\cite{fu2008superconducting},
	this subject has emerged as a highly active research area in the field of condensed matter physics. Extensive efforts have focused on realizing MZMs 
	in diverse platforms such as semiconductor-superconductor heterostructures~\cite{frolov2020topological,lutchyn2018majorana,mourik2012signatures,oreg2010helical} and magnetic impurity Shiba chains~\cite{teixeira2020enhanced,mashkoori2019majorana,andolina2017topological,reis2014self,poyhonen2014majorana,sau2013bound,braunecker2013interplay,nadj2013proposal,pientka2013topological}. In particular, the elegant idea of heterostructures comprising of a one-dimensional (1D) nanowire with Rashba spin-orbit coupling (SOC), a time-reversal symmetry (TRS) breaking component (\eg Zeeman field) and proximity-induced isotropic ($s$-wave) superconductivity has been theoretically proposed as a promising route to observe topological superconductivity (TSC) hosting MZMs~\cite{oreg2010helical,lutchyn2018majorana,ghorashi2024altermagnetic,prada2020andreev,mondal2025distinguishing,wu2020plane} and has also garnered much experimental efforts~\cite{mourik2012signatures,albrecht2016exponential,nadj2014observation,shabani2016two,schiela2024progress,zhu2021majorana}. 
	
    From this perspective, $D$-dimensional topological superconductors can host $n$-dimensional Majorana boundary modes ($n=D-d$, where $D$ is the system dimension and $d$ is the topological order). Following this definition, first-order (FOTSC) and second-order topological superconductivity (SOTSC) manifest as 1D dispersive/flat Majorana edge modes (MEMs) or zero-dimensional (0D) Majorana corner modes (MCMs). 
    In recent times, a quantum spin Hall insulator (QSHI) in the presence of in-plane Zeeman field and proximity induced superconductivity has been shown to host 
    MCMs in two dimensions (2D) and hinge modes in three-dimensions (3D)~\cite{wu2020plane}. 
    Furthermore, static magnetic ordering such as in noncollinear spin texture based 2D heterostructures~\cite{nakosai2013two,yan2018majorana,zlotnikov2021aspects,subhadarshini2025engineering,chatterjee2024second,chatterjee2024topological,subhadarshini2024multiple} have been shown to exhibit SOTSC~\cite{chatterjee2024second,subhadarshini2025engineering}, where modulation of the pitch vector 
    leads to an effective SOC that plays a pivotal role in the topological phase transition.
	
    In recent times, altermagnets (AMs) have picked up tremendous research attention in the scientific community. These are even-parity, compensated with collinear magnetic order arising from crystal symmetry and exhibit spin-split bands~\cite{song2025altermagnets,tamang2024newly,gomonay2024structure,vsmejkal2022emerging}. Heterostructures based on such AMs have become a rich playground that exhibit TSC of various orders in all three dimensions~\cite{ghorashi2024altermagnetic,mondal2025distinguishing,alam2026proximity,yi2026majorana,hodge2026altermagnet,subhadarshini2026current}. 
    In similar spirit, recent theoretical 
    demonstrations have put forward another kind of unconventional magnetism designated as $p$-wave magnets ($p$WMs), that exhibits odd-parity spin-split band structure and are experimentally accessible materials~\cite{hellenes2023p,brekke2024minimal,sim2026quantum,ezawa2024topological,song2025electrical,yamada2025metallic}. Such $p$WMs also possess net zero magnetization but host noncollinear magnetic order. They have broken parity ($\mathcal{P}$) and $\mathcal{PT}$ symmetries, but preserve a composite TRS $\mathcal{T}$ that results in a net zero magnetization~\cite{brekke2024minimal}. Intrinsic superconducting properties of such unconventional $p$WMs
    exhibit weak first-order TSC and superconducting diode effect~\cite{pal2026emergent,fukaya2026p,patra2026floquet}.
   However, heterostructures based on $p$WMs has not been considered so far to engineer hierarchy (both first and second order) of TSC phases in 2D. 
   In this letter, we pose the following intriguing questions: (i) how does the edge states of a quantum spin Hall insulator (QSHI) modify in presence of the unconventional $p$-wave magnetic order and proximity-induced regular $s$-wave superconducting pairing? (ii) can such a platform host second order topological superconducting (SOTSC) phase anchoring Majorana corner modes (MCMs)? 
   (iii) can this phase transition be determined analytically via low-energy edge theory? To answer these questions, we propose a model based on a hybrid platform consisting of a QSHI, $p$WM and conventional $s$-wave superconductor (SC), where, 
   this lattice model does not rely on any Zeeman field or relativistic SOC induced by heavy elements but also preserves TRS $\mathcal{T}$. We show that such heterostructure utilizes the intrinsic SOC of QSHI mixed with the anisotropic spin-dependent hopping strength of the $p$WM 
   in the presence of $s$-wave superconductivity to obtain 
   both edge and corner-localized Majorana modes in this system. 

 	\begin{figure}[h]
		\includegraphics[width=0.49\textwidth]{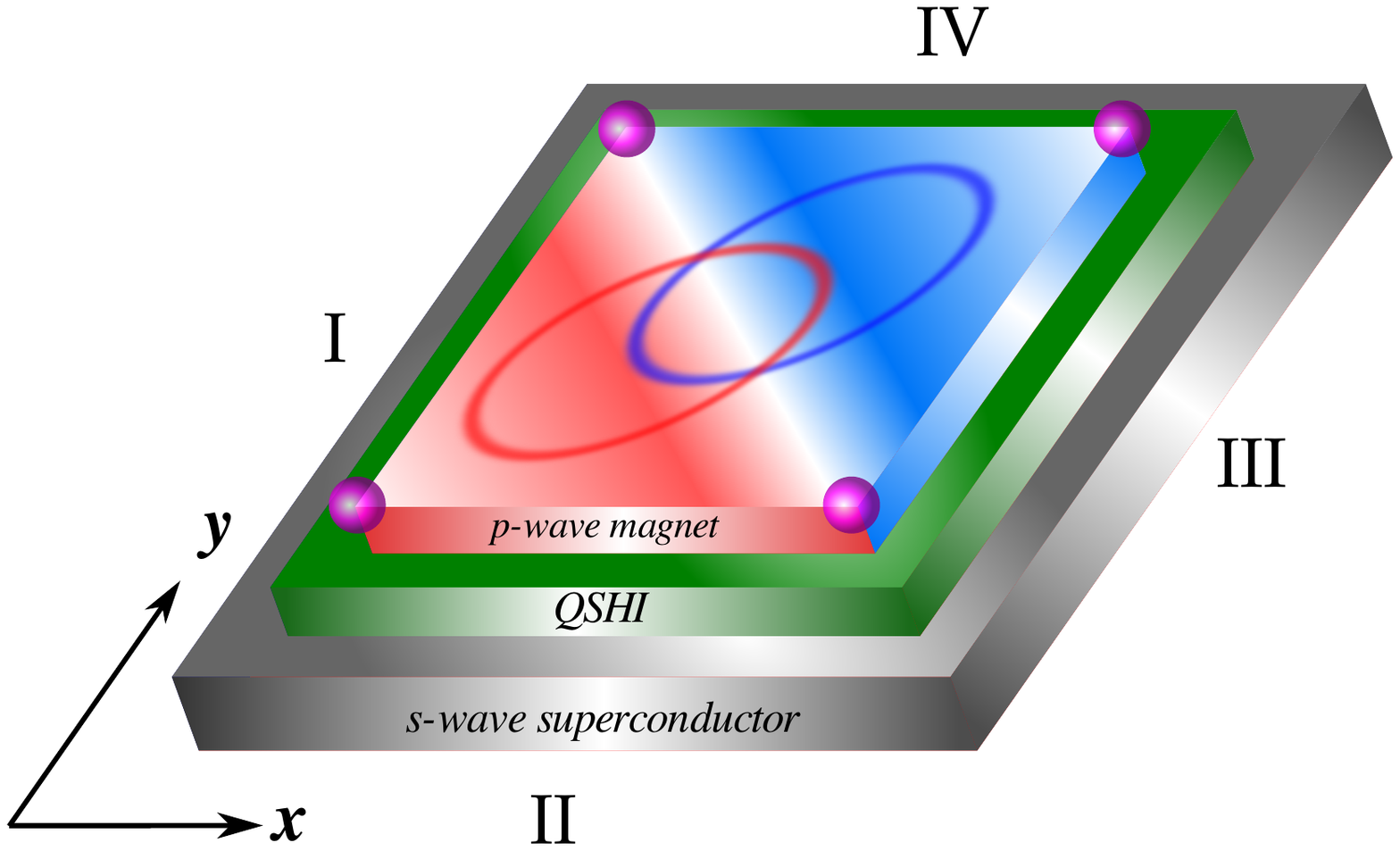}
		\caption{Schematic of our heterostructure is demonstrated comprising of a conventional $s$-wave SC, a QSHI and a non-collinear $p$WM. In the SOTSC phase, 
		the system hosts zero-energy MCMs, localized at each corner of the 2D domain. The four edges are denoted by I, II, III, IV.}
		\label{Fig1}
	\end{figure}
{\it{\textcolor{blue}{Model Hamiltonian}}}:-
    We consider the heterostructure presented in Fig.~\ref{Fig1} which consists of a non-collinear $p$WM, a 2D topological insulator (QSHI) and $s$-wave SC. 
    This composite system is modeled by employing the Bogoliubov-de Gennes (BdG) Hamiltonian in momentum-space, 
    given by~\cite{tkachov2015topological,chung2011conductance,qi2010chiral,wu2021anderson},
	\begin{equation}
		H_{\rm BdG}(\boldsymbol{k}) =\begin{bmatrix}
		H_0(\boldsymbol{k})) &  \Delta \\
		-\Delta^* & -H^*_0(-\boldsymbol{k}))
		\end{bmatrix}\ ,
		\label{Eq1}
	\end{equation}
	where, $ H_0(\boldsymbol{k}) = H_{{\rm{BHZ}}}(\boldsymbol{k}) + H_{p\rm WM}(\boldsymbol{k})$. Here, $H_{{\rm{BHZ}}}$ corresponds to the Bernevig-Hughes-Zhang (BHZ) Hamiltonian~\cite{ghosh2024generation,bernevig2006quantum} describing the QSHI. The latter is given as
	\begin{equation}
	  \begin{split}
		H_{\rm BHZ}(\boldsymbol{k}) = [m_0-t(\cos{k_x}+\cos{k_y})]~\tau_z\sigma_0 &+ \\ \lambda_x \sin{k_x}~\tau_x\sigma_z + 	\lambda_y \sin{k_y}~\tau_y\sigma_0\ .
 	 \end{split}
	\end{equation}
	The unconventional $p$-wave magnetism is incorporated using the minimal model Hamiltonian~\cite{brekke2024minimal}
	\begin{align}
		H_{p\rm WM}(\boldsymbol{k}) &= (\alpha_x \sin{k_x}+\alpha_y \sin{k_y}) \tau_0\sigma_z + J_{sd} \tau_z\sigma_x\ ,
	\end{align}
	The proximity-induced $s$-wave superconducting pairing is given by~\cite{bruus2004many,chung2011conductance,qi2010chiral,wu2020plane,wu2021anderson}
	\begin{align}
		\Delta &= \Delta_0 ~ ( i \tau_0  \sigma_y) \nonumber\\
		&= \Delta_0 ~ ( 
		c_{\boldsymbol{k}\uparrow}^\dagger c_{\boldsymbol{-k}\downarrow}^\dagger 
		- c_{\boldsymbol{k}\downarrow}^\dagger c_{\boldsymbol{-k}\uparrow}^\dagger
		+ c_{\boldsymbol{-k}\downarrow} c_{\boldsymbol{k}\uparrow}
		- c_{\boldsymbol{-k}\uparrow} c_{\boldsymbol{k}\downarrow})\ ,
	\end{align}
	The Nambu-Gorkov basis used here is denoted as 
	$\Psi_{\boldsymbol{k}} = \Big( c_{\boldsymbol{k}A\uparrow} , c_{\boldsymbol{k}A\downarrow} , c_{\boldsymbol{k}B\uparrow} , c_{\boldsymbol{k}B\downarrow} , c_{\boldsymbol{-k}A\uparrow}^\dagger , c_{\boldsymbol{-k}A\downarrow}^\dagger , c_{\boldsymbol{-k}B\uparrow}^\dagger , c_{\boldsymbol{-k}B\downarrow}^\dagger \Big)^T$. Here, the particle-hole, orbital and spin spaces are denoted by the Pauli matrices $\boldsymbol{\rho},\boldsymbol{\tau}$	and $\boldsymbol{\sigma}$, respectively. Also, $m_0,t,\lambda_{x,y}$ represent the staggered Dirac mass, the nearest-neighbor hopping amplitude and the strength of the intrinsic SOC along $x$ and $y$-directions, respectively. The isotropic $s$-wave superconducting pairing amplitude is represented by $\Delta_0$. The non-collinear nature of the $p$WM is adapted through the spin-split hoppings $\alpha_{x,y}$ and the isotropic $sd$ coupling between the localized $s$-electrons with itinerant $d$-electrons with strength given by 
	$ J_{sd}$. This composite system breaks the parity symmetry $\mathcal{P}=\rho_0\tau_z\sigma_{0}$, but preserves a combined TRS: $\mathcal{T}=\rho_{0}\tau_0\sigma_x \mathcal{K}$. Therefore, $\mathcal{T}^{-1}H_{\rm BdG}(\boldsymbol{k})\mathcal{T}=H_{\rm BdG}(-\boldsymbol{k})$. In all our numerical analysis, we choose $t=1,~\Delta_0=0.5$ and $\lambda_{x,y}=1$, unless stated otherwise. All other model parameters are scaled by the
	hopping amplitude $t$. 
	

{\it{\textcolor{blue}{Emergence of hierarchy of TSC phases}}}:-
	QSHI is well-known to exhibit first-order band topology, manifested in the form of linearly dispersive 1D helical edge modes~\cite{bernevig2006quantum,tkachov2015topological,shen2012topological,bernevig2013topological,imura2010zigzag,ferreira2022engineering} at the boundary. These modes can be trivially gapped out in presence of proximity induced $s$-wave superconducting pairing~\cite{fu2008superconducting,hart2014induced}. When $\Delta_0=0$, QSHI coupled with $p$WM also exhibits boundary modes; but because the latter breaks the $C_4$ symmetry, the dispersion of the edge modes along one direction becomes gapped whereas in the other direction it remains gapless. Also, the gapless dispersion becomes comparatively flat due to the competition between $\alpha_{x,y}$ and $\lambda_{x,y}$. 
	Very recently, $p$WMs are shown to exhibit FOTSC hosting flat Majorana edge modes~\cite{pal2026emergent} similar to general non-collinear spin texture with broken TRS~\cite{chatterjee2024topological}. This feature indicates towards the prospect of a hybrid system comprising of QSHI, $p$-wave magnetism and $s$-wave superconductivity as considered here in 
	Fig.~\ref{Fig1}. This may lead to a mass gap inversion at the junction of adjacent edges, thereby giving
	rise to SOTSC phase hosting MCMs (see latter text for discussion). 
	
	\begin{figure}[t]\centering
		\includegraphics[width=0.49\textwidth]{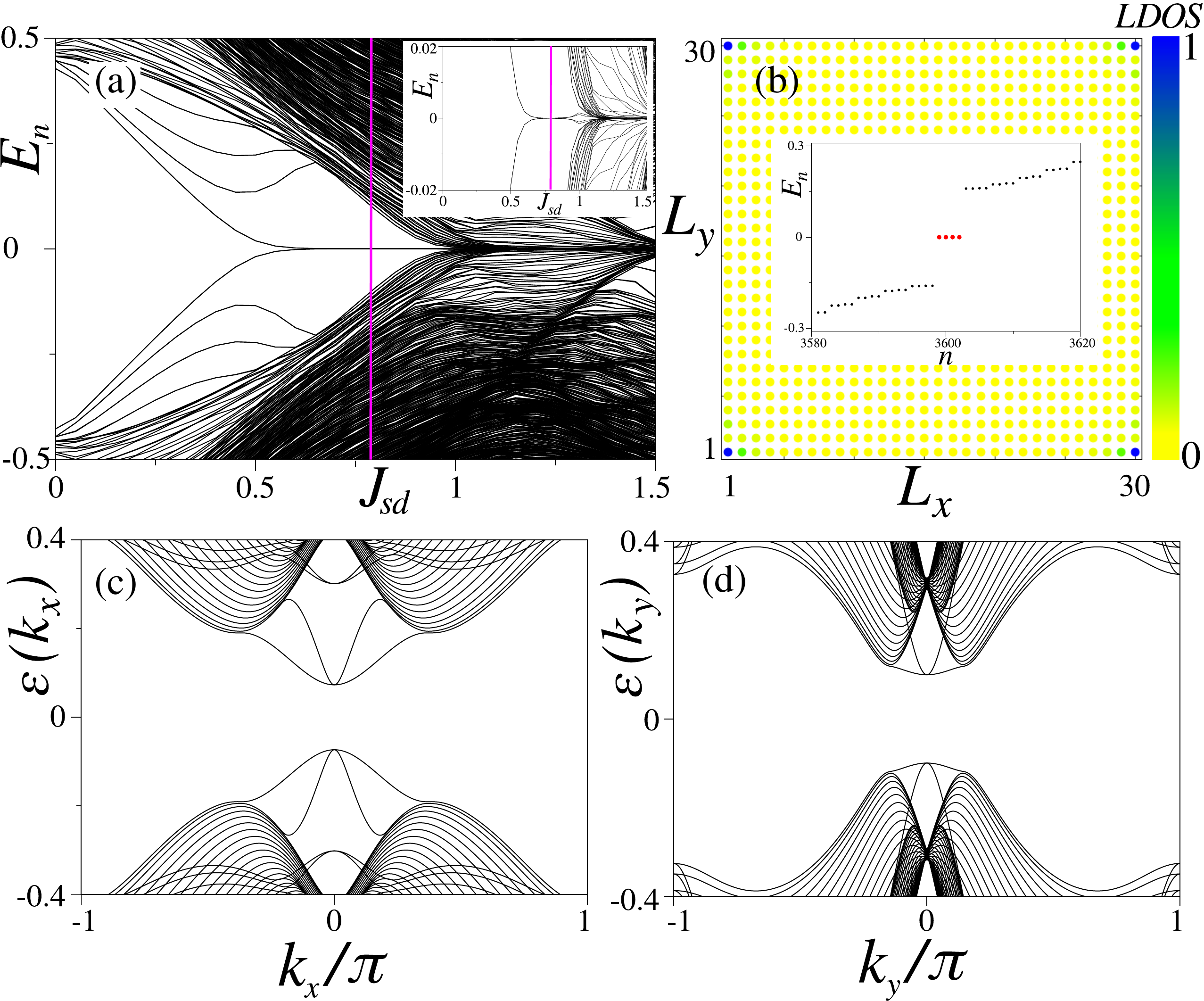}
		\caption{(a) Eigenvalue spectra of $H_{\rm BdG}$  (Eq.~(\ref{Eq1})) is displayed as function of $J_{sd}$ 
		employing OBC. For a given $J_{sd}$, marked by the magenta line, the system exhibits SOTSC phase with the inset highlighting the zoomed in version of that region. The corresponding LDOS at $E=0$ is presented in panel (b), along with the eigenvalue spectrum as a function of state index $n$ in the inset. The four MCMs are indicated by red dots, and located well inside the bulk gap. At SOTSC phase, the edge modes along both $x$ and $y$ directions are gapped out and are presented in panels (c) and (d), respectively.  Here, we consider a finite size system with $30\times30$ lattice sites and
       the model parameters are chosen as: $ m_0=1,J_{sd}=0.8,\alpha_{x,y}=0.5$. The energy scale for all model parameters are set with respect to the hopping amplitude $t$.}
		\label{Fig2}
	\end{figure}	
	We begin by analyzing the variation of the eigenvalue spectrum, corresponding to the heterostructure, 
	as a function of $J_{sd}$, subject to 2D open-boundary condition (OBC).  The result is shown in 
	Fig.~\ref{Fig2}(a). Initially, for small value of $J_{sd}$, the system behaves similar to a trivial SC and the spectrum remains trivially gapped. As we increase $J_{sd}$, the edge-modes gradually come closer to $E=0$, and in the region $0.55 < J_{sd}<1$ the system makes a transition 
	to the SOTSC phase where the system hosts localized MCMs as shown in Fig.~\ref{Fig2}(b) via the eigenvalue spectrum and local density of states (LDOS). 
	Moreover, the edge modes are gapped out along both $x$ and $y$ directions
	as depicted in Figs.~\ref{Fig2}(c)-(d). Beyond $J_{sd}\approx 1$, the system becomes trivially gapped for a very small range of $J_{sd}$ and then remains gapless (see the inset of Fig.~\ref{Fig2}(a)), thereby leading to the onset of FOTSC. In this regime, edge-modes appear along the $x$-direction and the $y$-direction remains gapped. Such Majorana edge modes are flat (non-dispersive ) and appear within a weak FOTSC phase as we discuss later. 
    In light of these discussions, our model Hamiltonian exhibits both first and second-order band topology 
    that can be tuned with the parameters of the non-collinear $p$WM.  

{\it{\textcolor{blue}{Characterization of FOTSC phase}}}:-
In addition to TRS, the Hamiltonian in Eq.~(\ref{Eq1}) also preserves the charge-conjugation symmetry $\mathcal{C}=\rho_x\tau_0\sigma_0 \mathcal{K}$ and the chiral symmetry $\mathcal{S}=\rho_0\tau_x\sigma_x$ \ie  $\mathcal{C}^{-1}H_{\rm BdG}(\boldsymbol{k})\mathcal{C}=-H_{\rm BdG}(-\boldsymbol{k})$ and $\mathcal{S}^{-1}H_{\rm BdG}(\boldsymbol{k})\mathcal{S}=-H_{\rm BdG}(\boldsymbol{k})$ respectively. Thereby this system falls into the 
BDI class~\cite{shen2012topological,altland1997nonstandard,schnyder2008classification} where a 1D $Z$ invariant can be defined~\cite{pal2026emergent}. 
Therefore the Hamiltonian can be expressed in the anti-diagonal form $\mathcal{H}=\left(\begin{smallmatrix} 0 & q(\boldsymbol{k})\\ q^\dagger(\boldsymbol{k}) & 0 \end{smallmatrix}\right)$ (see the SM for details~\cite{suppmat}). This leads to the definition for the 1D Winding number as
	\begin{equation}		
       \mathcal{N}_y(k_x) = \frac{i}{2\pi} \int_{-\pi}^{\pi} dk_y ~\rm{Tr} \left[{q^{-1}(\boldsymbol{k}}) \frac{\partial q(\boldsymbol{k})}{\partial k_y}\right]\ .
    \end{equation}
	
\begin{figure}
		\includegraphics[width=0.5\textwidth]{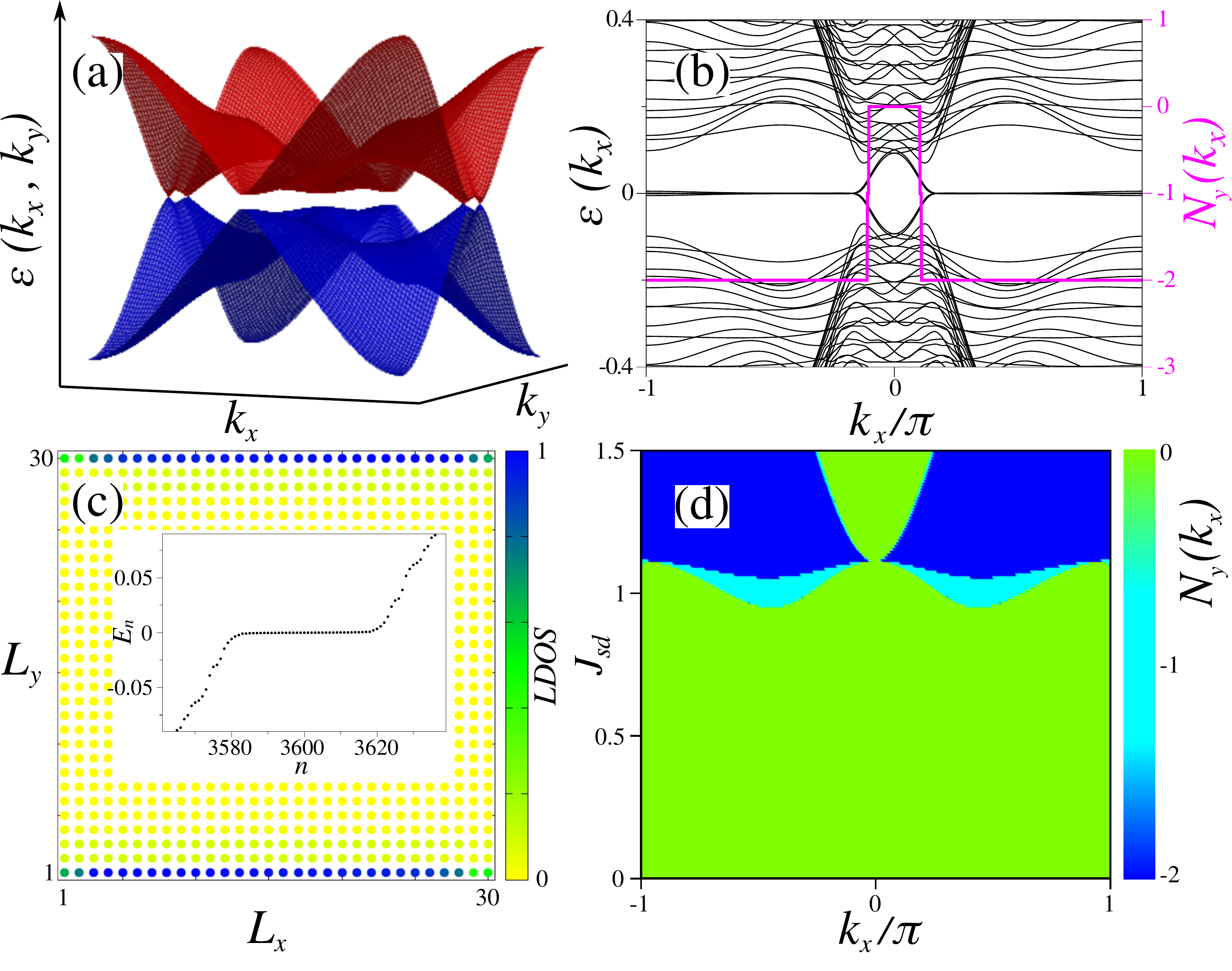}
		\caption{(a) Bulk spectrum in the FOTSC phase is depicted in $k_{x}$-$k_{y}$ plane exhibiting four gapless points. (b) The corresponding spectrum considering a ribbon geimetry is shown as a function of $k_x$ manifesting the MFEMs, characterized by the winding number $\mathcal{N}_y$ (depicted in magenta). (c) LDOS spectra (at $E=0$) is displayed in $L_{x}-L_{y}$ plane considering a finite size system with $30\times30$ lattice sites. The inset exhibits the eigenvalue
		spectrum as a function of the eigenvalue index (employing OBC along both $x$ and $y$ directions) manifesting the MFEMs. (d) Variation of $\mathcal{N}_y$ 
		is shown in the plane of $J_{sd}$ and $k_x$. For panels (a,b,c), the model parameters are chosen as $ m_0=1, J_{sd}=1.2,\alpha_{x,y}=0.5$. For panel (d), $J_{sd}$ is varied and other model parameters remain same as mentioned above.}
		\label{Fig3}
	\end{figure}

	In the FOTSC regime, the bulk spectrum is gapped except at four isolated gapless points as shown in Fig.~\ref{Fig3}(a). Such gapless phase hosts Majorana flat edge modes (MFEMs) that appear at zero-energy in the band dispersion corresponding to the ribbon geometry (see Fig.~\ref{Fig3}(b)). Emergence of MFEMs has been reported earlier in literature in other context~\cite{chatterjee2024topological,subhadarshini2024multiple}. 
	Such gapless FOTSC phase can be characterized by the 1D winding number $\mathcal{N}_y=-2$, thereby topological in nature (see Fig.~\ref{Fig3}(b)). The LDOS 
	distribution at $E=0$ is depicted in Fig.~\ref{Fig3}(c) considering finite sheet geometry. 
	This exhibits the edge localization of MFEMs along the $x$-direction as a consequence of the broken $C_4$ symmetry in $p$WMs. As discussed earlier, the FOTSC phase emerges for $J_{sd}>1$ (see Fig.~\ref{Fig2}(a)) and characterized by $\mathcal{N}_y(k_{x})$ as displayed by Fig.~\ref{Fig3}(d) in the phase space of $(J_{sd},k_x)$. 

\begin{figure}
		\includegraphics[width=0.47\textwidth]{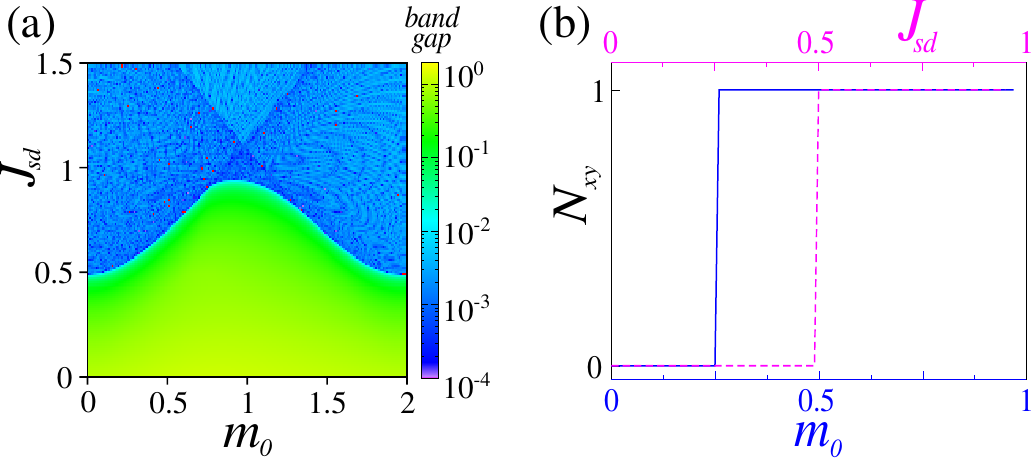}
		\caption{(a) Bulk band gap is shown in the $(J_{sd} - m_0)$ plane. (b) $\mathcal{N}_{xy}$ is depicted 
		as a function of $J_{sd}$ and $m_0$, indicated by the magenta and blue color lines, respectively. 
		Here, $\mathcal{N}_{xy}$ versus $J_{sd}$ is calculated at $m_0=1$. For $\mathcal{N}_{xy}$ versus $m_0$, the $J_{sd}$ is fixed at $0.5$. Other model parameters are chosen as $\alpha_{x,y}=0.5, \lambda_{x,y}=1.0, \Delta_0=0.5$.}
		\label{Fig4}
	\end{figure}

{\it{\textcolor{blue}{Characterization of SOTSC phase}}}:-
	In order to characterize the SOTSC phase, a higher-order topological invariant known as quadrupolar winding number ($\mathcal{N}_{xy}$) can be defined owing to the chiral symmetry of the system~\cite{benalcazar2022chiral,pal2025multi}. This follows as
	\begin{equation}
		\mathcal{N}_{xy} = \frac{1}{2\pi i} \rm{Tr} \Big[\log \big(\bar{Q}_{xy}^A \bar{Q}_{xy}^{B\dagger}\big) \Big] \in \mathbb{Z}\ ,
		\label{Qxy}
	\end{equation}
   where, $\bar{Q}_{xy}^S=U_S^\dagger Q_{xy}^S U_S$ is the orbital/sublattice quadrupole moment projected into the subspace $U_S$ for $S=A,B$ sublattices. 
   Here, $Q_{xy}^S =  \sum_{\boldsymbol{R}} |\boldsymbol{R}\rangle \exp\left(-i \frac{2\pi xy}{L_x L_y} \right) \langle \boldsymbol{R}| $ is the sublattice quadrupole moment operator and $U_S$ denotes the matrix formed by the eigenvector of the chiral operator $\mathcal{S}$ (see SM for details~\cite{suppmat}).
   We compute the bulk band gap (energy difference benween highest valence and lowest conduction band) 
   in $(J_{sd}-m_0)$ plane using periodic boundary condition and calculate the quadrupolar winding number $\mathcal{N}_{xy}$  employing Eq.~(\ref{Qxy}). The corresponding behavior is shown in Fig.~\ref{Fig4}. As seen previously from Fig.~\ref{Fig2}(a), the SOTSC phase appears in the $(J_{sd}-m_0)$ plane  when the bulk is significantly gapped (see Fig.~\ref{Fig4}(a)). Within the gapped topological regime, the quadrupolar winding number $\mathcal{N}_{xy}$ takes quantized non-zero integer values, and zero in the trivially gapped regime. From Fig.~\ref{Fig4}(b) (magneta line), it is evident that for a fixed $m_0$ when the bulk is significantly gapped (see Fig.~\ref{Fig4}(a)), the SOTSC phase is characterized by $\mathcal{N}_{xy}=1$. This phase anchors four zero-energy localized MCMs. Similarly, upon keeping $J_{sd}$ fixed and varying $m_0$, the system transits from trivial to SOTSC phase which is characterized by $\mathcal{N}_{xy}=1$ along the blue line as shown in Fig.~\ref{Fig4}(b).

    
{\it{\textcolor{blue}{Low-energy edge theory}}}:-
	Here, we formulate an effective low-energy edge theory for the 2D heterostructure shown in Fig.~\ref{Fig1}. Evaluating near the $\Gamma$ point, the approximation  $\sin k_{x,y} \approx k_{x,y}$ and $\cos k_{x,y} \approx 1-\frac{k^2_{x,y}}{2}$ holds true, which leads to an effective low-energy Hamiltonian:
	\begin{align}
		H_{\rm eff} = &\Big[m_0-t\Big(1-\frac{k_x^2}{2}\Big)-t\Big(1-\frac{k_y^2}{2}\Big)\Big]\rho_z\tau_z\sigma_0 \nonumber \\
		&~+~ \lambda_x k_x ~\rho_0\tau_x\sigma_z ~+~ \lambda_y k_y 	~\rho_z\tau_y\sigma_0 \nonumber \\ &~+~ \big(\alpha_x k_x + \alpha_y k_y\big)~\rho_0\tau_0\sigma_z + J_{sd} ~\rho_z\tau_z\sigma_x \nonumber \\ &~-~ \Delta_0 ~\rho_y\tau_0\sigma_y\ ,
		\label{Heff}
	\end{align}
    Let us consider the edge-I along $y$-direction of the heterostructure. For this case, $k_x$ is substituted by $-i\partial_x$, and neglecting the $k_y^2$ terms, $H_{\rm eff}$ can be split into an exact part $H_e$ and a perturbation $H_p$ as,
    \begin{align}
    	H_e &~=~  \Big(m - \frac{t}{2} \partial_x^2\Big)~ \rho_z\tau_z\sigma_0 ~-~ i \lambda_x \partial_x ~ \rho_0\tau_x\sigma_z\ , \\
    	H_p &~=~ \lambda_y k_y ~\rho_z\tau_y\sigma_0 ~+~ \Big(-i \alpha_x \partial_x + \alpha_y k_y  \Big) \nonumber ~\rho_0\tau_0\sigma_z \\ &~+~ J_{sd} ~\rho_z\tau_z\sigma_x  ~-~ \Delta_0 \rho_y\tau_0\sigma_y\ ,
    	\label{Heff2}
    \end{align}
    The effective Dirac mass becomes $m = m_0 - 2t$. Subject to the 
    Fu-Kane criterion~\cite{fu2007topological} $m<0$, $H_e$ can be solved exactly and the corresponding solutions are used to obtain the matrix elements of $H_p$. The corresponding calculational details are elaborated in the SM~\cite{suppmat}. Therefore, the Hamiltonian for edge-I is given by:
    \begin{equation}
    	H_{\rm edge}^{\rm I} = (\lambda_y-\alpha_y) k_y ~\tau_0\sigma_z + J_{sd} ~\tau_z \sigma_z - \Delta_0 ~\tau_y\sigma_y\ ,
    \end{equation}
    This resembles to a 1D Dirac equation. Following the similar procedure, the Hamiltonian for edge-II can be obtained as:
    \begin{equation}
    	H_{\rm edge}^{\rm II} = -(\lambda_x+\alpha_x) k_x ~\tau_0\sigma_z - \Delta_0 ~\tau_y\sigma_y\ ,
    \end{equation}
    According to the Jackiw-Rebbi condition~\cite{shen2012topological,jackiw1976solitons}, localized modes appear at the interface between Dirac masses with opposite signs. Investigation of the edge Hamiltonians reveals a mass inversion when $J_{sd}>\Delta_0$ (for edge I while edge II always remains massive with same sign), and thus a MCM solution is obtained at the junction of two adjacent edges.
    
\begin{figure}
    	\includegraphics[width=0.49\textwidth]{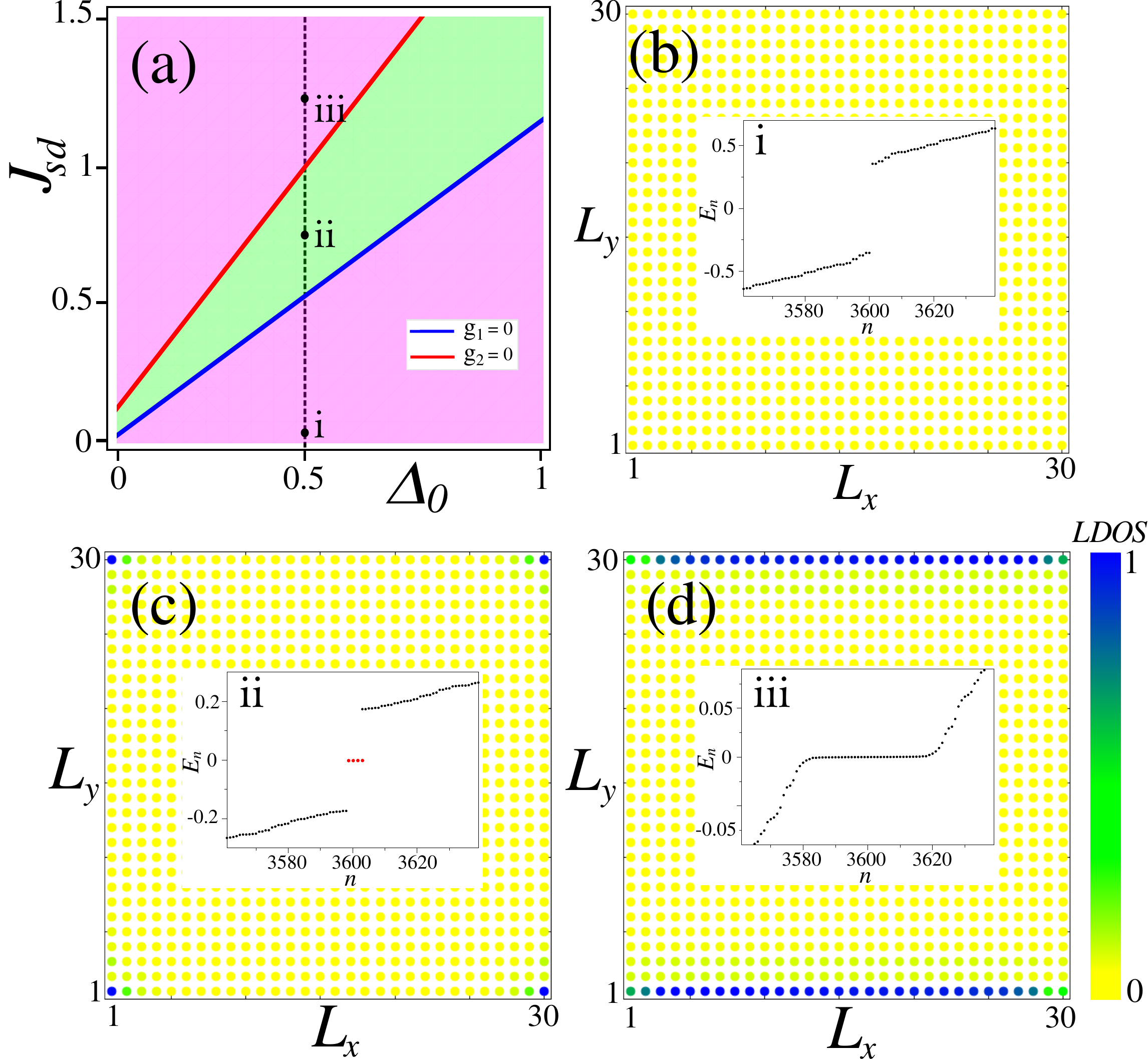}
    	\caption{(a) The mass (band) gap profile (Eq.~(\ref{mg})) is illustrated in $J_{sd}-\Delta_0$ plane. Here, $g_1$ and $g_2$ correspond to the mass gaps for $H_{\rm edge}^{\rm I}$ and $H_{\rm edge}^{\rm II}$, respectively. The green region indicates the validity of Jackiw-Rebbi condition and the magenta region imply otherwise. 
    	In panels (b,c,d), we depict the zero-energy LDOS distribution and eigenvalue spectrum (inset) obtained employing OBC corresponding to the points
    	(i,ii,iii) of panel (a) respectively. 
    	The other model parameters are considered as $ m_0=1,\alpha_{x,y}=0.5,\lambda_{x,y}=1$.} 
    	\label{Fig5}
    \end{figure}

    To gain deeper insight into the appearance of hierarchy of topological phases, we incorporate $\alpha_{x,y}$ in $H_0$ for the respective edges, and obtain a more precise Jackiw-Rebbi condition
    given as,
    \begin{equation}
    	(J_{sd}\sqrt{1-\eta_x^2}-\Delta_0)(J_{sd}~\eta_y-\Delta_0) < 0\ ,
    	\label{mg}
    \end{equation}
    where, $\eta_{x,y}=\frac{\alpha_{x,y}}{\lambda_{x,y}}$. The mass gap profile from the above equation is displayed in Fig.~\ref{Fig5}(a). 
	At a fixed $\Delta_0$, variation of $J_{sd}$ exhibits transitions of the heterostructure through three different phases. At low $J_{sd}$, the system is in the trivial phase without any zero-energy states within the bulk gap. The corresponding eigenvalue spectrum and LDS are shown in Fig.~\ref{Fig5}(b) by choosing a point ``i" from that region. With gradual increase of $J_{sd}$, the system makes a transition from trivial to SOTSC phase at intermediate $J_{sd}$. This is indicated by the presence of four MCMs in the eigenvalue spectrum and their corner localization in LDOS (see Fig.~\ref{Fig5}(c)). The latter refers to a point
	``ii" chosen from that region in $J_{sd}-\Delta_0$ plane. In the high $J_{sd}$ limit, the bulk gap becomes significantly narrow (gapless) hosting zero-energy MFEMs as shown in Fig.~\ref{Fig5}(d) via eigenvalue spectrum and LDOS, corresponding to a point ``iii" chosen from that region in $J_{sd}-\Delta_0$ plane. The two solid lines (denoted by red and blue) correspond to the massless ($g_{1}=g_{2}=0$) lines \ie topological phase boundary.
	

{\it{\textcolor{blue}{Bulk pairing analysis}}}:-
	Following the discussion on emergence of hierarchy of topological phases indicated by the appearance of corresponding Majorana modes, here we present the effective superconducting bulk pairings arising from the interplay of the QSHI, $p$-wave magnetic order and $s$-wave superconductivity. Here, we employ the effective Hamiltonian described in Eq.~(\ref{Heff}) and utilizing an unitary duality transformation~\cite{sato2009non,chatterjee2024second}, a dual Hamiltonian $H_D$ can be obtained of the form:
	 \begin{gather}
		H_D = (U_D)^\dagger H_{\rm eff} U_D \nonumber \\
		~~~~= \begin{pmatrix}
			\epsilon_D & \Delta_D \\ \Delta_D^\dagger & -\epsilon_D
		\end{pmatrix}\ ,                                                                               
	\end{gather}
    where, $\epsilon_D = \Delta_0 ~\tau_0\sigma_0 + J_{sd} ~\tau_x\sigma_x$ and $\Delta_D = \Delta_s + \Delta_{p_x+p_y} + \Delta_{p_x+ip_y}$. Here,
    \begin{align}
    	\Delta_s &= -i(m_0 - 2t) \tau_z\sigma_y\ , \nonumber \\
    	\Delta_{p_x+p_y} &= -(\alpha_x k_x + \alpha_y k_y) ~\tau_0\sigma_x\ , \nonumber \\
    	\Delta_{p_x+ip_y} &= -(\lambda_x k_x ~\tau_x\sigma_x + i \lambda_y k_y ~\tau_y\sigma_y) \ .\ 
    	\label{Delta}
    \end{align}

    Note that, the minus signs in Eqs.~(\ref{Delta}) appear due to the choice of basis (we refer to SM for further details~\cite{suppmat}). We observe that two types of non-isotropic pairing emerge depending on two different parameters. The $\Delta_{p_x+p_y}$ pairing depends on the intrinsic SOC of the QSHI ($\lambda_{x,y}$), whereas the $\Delta_{p_x+i p_y}$ is proportional to the SOC like term induced by the spin-split hopping of the $p$WM ($\alpha_{x,y}$). As the bulk $(p_x+p_y)$ pairing is known to exhibit MFEMs~\cite{nakosai2013two,chatterjee2024topological,wang2017topological,zhang2019majorana} in contrast to $(p_x+ip_y)$ pairing which exhibits dispersive Majorana edge modes~\cite {nakosai2013two,hu2019chiral}, the nature of edge modes in our system also manifest the same. The relative magnitude of these bulk pairings can be modulated by the relative values of $\lambda_{x,y}$ and $\alpha_{x,y}$ at low-energy and near $\boldsymbol{k}\approx 0$. In the limit $\alpha_{x,y} \approx 0$, the $(p_x+ip_y)$ pairing dominates and all four dispersive edge modes are gapped giving rise to SOTSC. 
    On the other hand, $\lambda_{x,y} \approx 0$ results in flat edge modes that lead to gapless FOTSC hosting boundary modes at opposite edges~\cite{pal2026emergent,fukaya2026p} as $p$WM breaks $C_4$ but preserves $C_2$ symmetry. Hence, both $\lambda_{x,y}$ and $\alpha_{x,y}$ are required to 
    realize the reported hierarchy of TSC phases anchoring respective types of Majorana modes at the boundary.
    

{\it{\textcolor{blue}{Summary and discussions}}}:-
	To summarize, in this article, we investigate the phenomena of TSC in a 2D heterostructure comprising of a QSHI, an unconventional $p$WM and proximity-induced $s$-wave superconducting pairing. Our numerical analysis based on ribbon and sheet geometries reveal the emergence of both FOTSC and SOTSC phases as evident 
	from the eigenspectra and corresponding zero-energy LDOS distribution. Due to the presence of chiral symmetry, the system belongs to BDI class and consequently, a 1D Winding number $\mathcal{N}_y(k_{x})$ is employed 
	to characterize the gapless FOTSC phase hosting MFEMs. For the SOTSC phase, both the bulk as well as edge modes along both $x$ and $y$-direction are gapped 
	and quadrupolar winding number $\mathcal{N}_{xy}$ topologically characterizes this phase anchoring MCMs in finite 2D domain. The transitions between the three phases- trivial, SOTSC and FOTSC is shown to be achievable via the modulation of the parameters corresponding to the $p$WM. This has also been established
	analytically through an effective low-energy edge theory. Bulk pairing analysis reveals the interplay of emergent momentum-dependent superconducting pairings 
	$-(p_x+p_y)$ and $-(p_x+ip_y)$ that can be modulated by $\lambda_{x,y}$ and $\alpha_{x,y}$ respectively and are reponsible for the appearance of hierarchy of
	TSC phases in our hybrid system. 
	
	
    As far as experimental feasibility of our proposal is concerned, noncollinear $p$-wave magnetism has recently been proposed 
    in CeNiAsO~\cite{chakraborty2025highly} which exhibits a non-relativistic Edelstein effect that enables efficient spin-charge conversion. Another material displaying similar spin-polarized electronic bands is hexagonal Gd$_3$(Ru$_{1-\delta}$ Rh$_\delta$)$_4$
Al$_{12}~$\cite{yamada2025metallic} which hosts a spin-split band structure and coupled with broken inversion symmetry that results in anisotropic electrical conductivity. Notably, both these materials are metallic and do not rely on relativistic spin-orbit coupling from heavy metals. An insulating variant of $p$-wave magnet has been realized in NiI$_2$~\cite{song2025electrical} which exhibits spin-switching behavior. On the other hand, Nb(110) and Ta(110) are well-established conventional $s$-wave 
superconductors~\cite{odobesko2019preparation,schneider2021topological,heinrich2018single,beck2023systematic,beck2022structural}. In light of these developments, we propose a possible practical setup that involves positioning a QSHI between a top $p$-wave magnetic layer and a conventional $s$-wave superconducting substrate. To the best of our knowledge, such heterostructures based on odd-parity magnets has not yet been proposed in published literature. However, theoretical studies have demonstrated the coexistence of $p$-wave magnetism and 
FOTSC~\cite{sukhachov2025coexistence,nagae2025flat}. The latter can host MFEMs~\cite{pal2026emergent,patra2026floquet,luo2026hidden,ezawa2024topological}. In addition, we incorporate the QSHI emulating a HgTe/CdTe quantum well, whose interplay with the $p$WM leads to the realization of both MFEMs and MCMs in a single setup. In our model, the hopping strength is parametrized as $t=-B/{2a^2}$, where $B$ represents the effective kinetic term~\cite{bernevig2006quantum} and $a=1$ as the dimensionless lattice constant. Using $a\approx1$ nm and $B=-0.686$ eV.nm$^2$~\cite{li2009topological} one obtains $t\approx1.37$ eV. This serves as the reference energy scale to consistently scale the other model parameter values corresponding to Fig.~\ref{Fig2} as: $\lambda_x=\lambda_y\approx 1.37$ eV, $J_{sd}\approx 1.09$ eV, $\alpha_x=\alpha_y\approx 0.686$ eV and $\Delta_0\approx0.686$ eV. While these simulated values can effectively capture the low-energy physics presented here, we note that in real experimental scenarios these values may be substantially smaller.

   

{\it{\textcolor{blue}{Acknowledgments}}}:-
  K.R.D and A.S. acknowledge A. Pal and M. Subhadarshini for valuable discussions. K.R.D. and A.S. acknowledge the financial support from Department of Atomic Energy (DAE), Government of India 
  and the two Workstations provided by the Institute of Physics, Bhubaneswar from the DAE APEX Project 
  for numerical computations.
  
\textcolor{blue}{\textit{Data availibility statement:}-} The datasets generated and analyzed during the current study are available from the corresponding author upon reasonable request.

	\bibliography{refer}{}
	
	\normalsize\clearpage
	\begin{onecolumngrid}
		\begin{center}
			{\fontsize{12}{12}\selectfont
				\textbf{Supplemental Material for ``Hierarchy of topological superconductivity generated via heterostructures of unconventional $p$-wave magnets''\\[5mm]}}
			{\normalsize  Koushik R. Das\orcidA{} and $^{1,2}$ Arijit Saha\orcidB{},$^{1,2}$\\[1mm]}
			{\small $^1$\textit{Institute of Physics, Sachivalaya Marg, Bhubaneswar-751005, India}\\[0.5mm]}
			{\small $^2$\textit{Homi Bhabha National Institute, Training School Complex, Anushakti Nagar, Mumbai 400094, India}\\[0.5mm]}
		\end{center}
		
		\newcounter{defcounter}
		\setcounter{defcounter}{0}
		\setcounter{equation}{0}
		\renewcommand{\theequation}{S\arabic{equation}}
		\setcounter{figure}{0}
		\renewcommand{\thefigure}{S\arabic{figure}}
		\setcounter{page}{1}
		\pagenumbering{roman}
		
		\renewcommand{\thesection}{S\arabic{section}}

		The supplementary material (SM) contains the technical details corresponding to the theoretical analyses presented in the main text. In Sec.~\ref{Sec:I}, the derivation of the effective Hamiltonian is presented. 
		It is used in Sec.~\ref{Sec:II} to obtain an effective edge theory for adjacent edges and the Jackiw-Rebbi condition for the emergence of Majorana corner modes (MCMs). 
		The continuum effective Hamiltonian is also employed to gain insight into the effective superconducting pairings in the bulk as elaborated in Sec.~\ref{Sec:III}. Sec.~\ref{Sec:IV} is devoted to the discussion of the chiral symmetry of the Bogoliubov-de Gennes (BdG) Hamiltonian and the calculation of the quadrupolar winding  number.
		
		\section{DERIVATION OF the EFFECTIVE Low-energy Hamiltonian}\label{Sec:I}
		Here we present the derivation to obtain an effective low-energy Hamiltonian for the heterostructure discussed in the main paper. The BdG Hamiltonian in momentum-space is given by~\cite{tkachov2015topological,chung2011conductance,qi2010chiral,wu2021anderson}:
		\begin{equation}
			H_{\rm BdG} =\begin{bmatrix}
				H_0(k) &  \Delta \\
				-\Delta^* & -H^*_0(-k)
			\end{bmatrix}\ ,
		\end{equation}
		where, $ H_0(\boldsymbol{k}) = H_{\rm BHZ}(\boldsymbol{k}) + H_{p\rm {WM}}(\boldsymbol{k})$. 
		The quantum spin-Hall insulator (QSHI) is modeled using the Bernevig-Hughes-Zhang (BHZ) 
		Hamiltonian~\cite{ghosh2024generation,bernevig2006quantum}
		\begin{equation}
			H_{\rm BHZ}(\boldsymbol{k}) = [m_0-t(\cos{k_x}+\cos{k_y})] ~\tau_z\sigma_0 + \lambda_x \sin{k_x}~\tau_x\sigma_z + \lambda_y \sin{k_y}~\tau_y\sigma_0\ ,
		\end{equation}	
		The unconventional $p$-wave magnetism is incorporated using the minimal model~\cite{brekke2024minimal}
		\begin{align}
			H_{p\rm {WM}}(\boldsymbol{k}) &= (\alpha_x \sin{k_x}+\alpha_y \sin{k_y}) \tau_0\sigma_z + J_{sd} \tau_z\sigma_x\ ,
		\end{align}
		The proximity-induced $s$-wave superconducting pairing is given by~\cite{bruus2004many,chung2011conductance,qi2010chiral,wu2020plane,wu2021anderson}
		\begin{align}
			\Delta &= \Delta_0 ~ \tau_0 (i \sigma_y) \nonumber\\
			&= \Delta_0 ~ ( 
			c_{\boldsymbol{k}\uparrow}^\dagger c_{\boldsymbol{-k}\downarrow}^\dagger 
			- c_{\boldsymbol{k}\downarrow}^\dagger c_{\boldsymbol{-k}\uparrow}^\dagger
			+ c_{\boldsymbol{-k}\downarrow} c_{\boldsymbol{k}\uparrow}
			- c_{\boldsymbol{-k}\uparrow} c_{\boldsymbol{k}\downarrow})\ ,
		\end{align}
		The Nambu-Gorkov basis used here is denoted as
		\begin{equation}
			\Psi_{\boldsymbol{k}} = \Big( c_{\boldsymbol{k}A\uparrow} ~,~ c_{\boldsymbol{k}A\downarrow} ~,~ c_{\boldsymbol{k}B\uparrow} ~,~ c_{\boldsymbol{k}B\downarrow} ~,~ c_{\boldsymbol{-k}A\uparrow}^\dagger ~,~ c_{\boldsymbol{-k}A\downarrow}^\dagger ~,~ c_{\boldsymbol{-k}B\uparrow}^\dagger ~,~ c_{\boldsymbol{-k}B\downarrow}^\dagger \Big)^T \ ,
			\label{Psib}
		\end{equation}
		At low-energy, $(\sin{\boldsymbol{k}},\cos{\boldsymbol{k}})$ can be approximated to their respective Taylor expansions around ${\boldsymbol{k}}=0$ truncated to first-order, given by $\big({\boldsymbol{k}},(1-\frac{{\boldsymbol{k}}^2}{2})\big)$. Then the continuum, low-energy effective Hamiltonian can be written as
		\begin{align}
			H_{\rm eff} = &\left[ m_0-t\Big(1-\frac{k_x^2}{2}\Big)-t\Big(1-\frac{k_y^2}{2}\Big)   \right]\rho_z\tau_z\sigma_0 \nonumber \\
			&~+~ \lambda_x k_x ~\rho_0\tau_x\sigma_z ~+~ \lambda_y k_y 	~\rho_z\tau_y\sigma_0 \nonumber \\ &~+~ \big(\alpha_x k_x + \alpha_y k_y\big)~\rho_0\tau_0\sigma_z + J_{sd} ~\rho_z\tau_z\sigma_x ~-~ \Delta_0 ~\rho_y\tau_0\sigma_y\ .
			\label{Heff}
		\end{align}
		
		\section{LOW ENERGY EDGE THEORY}\label{Sec:II}
		In this section, we derive an effective low energy edge theory near $\boldsymbol{k}\approx0$, 
		similar 
		to~\cite{chatterjee2024second,subhadarshini2025engineering,yan2018majorana,wu2020plane,ghosh2021floquet}, using the effective Hamiltonian derived in Eq.~(\ref{Heff}). We here focus on adjacent edges designated as I (oriented along $y$-direction) and II (oriented along $x$-direction) as shown in Fig.~1 
		of the main text. Initially, we obtain the expressions by keeping the spin-split hopping terms $(\alpha_x,\alpha_y)$ of the $p$-wave magnet ($p$WM) in the perturbation part. Later, we 
		investigate the same by incorporating it in the unperturbed part and obtain the corresponding 
		Jackiw-Rebbi condition.
		
		\subsection{Hamiltonian for Edge-I}\label{Sec:II.A}
		In case of edge-I, we consider periodic boundary condition (PBC) along the $y$-direction and open boundary condition (OBC) along the $x$-direction. This leads to the substitution  $k_x \rightarrow -i\partial x$ in $H_{\rm eff}$, and the resulting Hamiltonian can be split into an exact part $H_e$ and the perturbation $H_p$. Then
		\begin{align}
			H_e &~=~  \Big(m - \frac{t}{2} \partial_x^2\Big)~ \rho_z\tau_z\sigma_0 ~-~ i \lambda_x \partial_x ~ \rho_0\tau_x\sigma_z\ , \\
			H_p &~=~ \lambda_y k_y ~\rho_z\tau_y\sigma_0 ~+~ \Big(-i \alpha_x \partial_x + \alpha_y k_y  \Big) ~\rho_0\tau_0\sigma_z ~+~ J_{sd} ~\rho_z\tau_z\sigma_x ~-~ \Delta_0 \rho_y\tau_0\sigma_y\ ,
		\end{align}
		where, $m=m_0-2t$ is the effective Dirac mass and the $k_y^2$ term is neglected in $H_p$. Assuming a zero-energy trial solution for $H_0$ as, 
		\begin{equation}
			\psi_\beta(x) = C e^{\gamma x} \chi_\beta\ ,
		\end{equation}
		the equation $H_e \psi_\beta(x) = 0$ can be written as
		\begin{equation}
			\det\Bigg[ \left(m - \frac{t}{2} \gamma^2 \right)~ \rho_z\tau_z\sigma_0 ~-~ i \lambda_x \gamma ~ \rho_0\tau_x\sigma_z \Bigg]=0\ .
		\end{equation}
		
		The solutions are obtained as
		\begin{eqnarray}
			\gamma &= \frac{-\lambda_x-\sqrt{2mt+\lambda_x^2}}{t}\ , 
			\label{S11}
		\end{eqnarray}
		\begin{eqnarray}
			\gamma &= \frac{\lambda_x-\sqrt{2mt+\lambda_x^2}}{t}\ , 
			\label{S12}
		\end{eqnarray}
		\begin{eqnarray}
			\gamma &= \frac{-\lambda_x+\sqrt{2mt+\lambda_x^2}}{t}\ , 
			\label{S13}
		\end{eqnarray}
		\begin{eqnarray}
			\gamma &= \frac{\lambda_x+\sqrt{2mt+\lambda_x^2}}{t} \ .
			\label{S14}
		\end{eqnarray}
		Subject to the boundary conditions: $\psi_\beta(0)=0 ~;~\psi_\beta(\infty)=0$, only Eq.~(\ref{S11}) and Eq.~(\ref{S13}) turn out to be the valid solutions. Incorporating these $\gamma$ expressions back into 
		the equation $H_e \psi_\beta(x) = 0$, the normalized eigenspinors can be obtained as
		\begin{equation}
			\chi_1=\frac{1}{\sqrt{2}}
			\begin{pmatrix}
				0\\0\\0\\0\\0\\i\\0\\1
			\end{pmatrix},
			\hspace{0.5cm}
			\chi_2=\frac{1}{\sqrt{2}}
			\begin{pmatrix}
				0\\0\\0\\0\\-i\\0\\1\\0
			\end{pmatrix},
			\hspace{0.5cm}
			\chi_3=\frac{1}{\sqrt{2}}
			\begin{pmatrix}
				0\\-i\\0\\1\\0\\0\\0\\0
			\end{pmatrix},
			\hspace{0.5cm}
			\chi_4=\frac{1}{\sqrt{2}}
			\begin{pmatrix}
				i\\0\\1\\0\\0\\0\\0\\0
			\end{pmatrix}\ .
		\end{equation}
		
		The Fu-Kane criterion~\cite{fu2007topological} dictates that $m<0$ in the topological regime \ie $m_0 < 2t$ for edge-theory to be applicable. Then,
		\begin{align}
			\gamma &= \kappa_1 \pm i \kappa_2\ , \\
			\kappa_1&=\frac{\lambda_x}{t} ~~;~~
			\kappa_2 =\sqrt{\frac{2|m|}{t}-\kappa_1^2}\ .  \nonumber  	
		\end{align}	
		Thus, the zero-energy trial solution is
		\begin{equation}
			\psi_\beta(x) = C e^{-\kappa_1 x} \sin(\kappa_2 x) ~ \chi_\beta  \hspace{1.5cm} (\beta=1,2,3,4)\ ,
		\end{equation}
		where the normalization constant $C = \frac{2}{\kappa_2} \sqrt{\kappa_1 (\kappa_1^2 + \kappa_2^2)}$. 
		The perturbation $H_p$ can then be expressed using ${\psi_\beta(x)}$ as basis states
		\begin{equation}
			(H_{\rm edge}^{\rm I})_{\beta_1,\beta_2} =  \int_{0}^{\infty} \langle \psi_{\beta_1}(x) | H_p | \psi_{\beta_2}(x)\rangle  ~ dx 
			\hspace{1cm} (\beta_1,\beta_2 = 1,2,3,4)\ ,
		\end{equation}
		Therefore, the Hamiltonian for edge-I can be written as
		\begin{equation}
			H_{\rm edge}^{\rm I} = (\lambda_y-\alpha_y) k_y ~\tau_0\sigma_z + J_{sd} ~\tau_z \sigma_z - \Delta_0 ~\tau_y\sigma_y\ .
		\end{equation}
		
		For the edge-III, identical procedure can be followed with the substitution $k_x \rightarrow -(-i\partial_x)$, and the corresponding edge Hamiltonian is then $H_{\rm edge}^{\rm III} = H_{\rm edge}^{\rm I}$. This is expected due to the underlying $C_2$ symmetry being preserved. 
		
		\subsection{Hamiltonian for Edge-II}\label{Sec:II.B}
		In case of edge-II, we consider PBC along the $x$-direction and OBC along the $y$-direction. This leads to the substitution  $k_y \rightarrow -i\partial y$ in $H_{\rm eff}$, and the resulting Hamiltonian can be split into an exact part $H_e$ and the perturbation $H_p$ as
		\begin{align}
			H_e &~=~  \Big(m - \frac{t}{2} \partial_y^2\Big)~ \rho_z\tau_z\sigma_0 ~-~ i \lambda_y \partial_y ~ \rho_z\tau_y\sigma_0\ , \\
			H_p &~=~ \lambda_x k_x ~\rho_0\tau_x\sigma_z ~+~ \Big(\alpha_x k_x - i \alpha_y \partial_y  \Big) ~\rho_0\tau_0\sigma_z ~+~ J_{sd} ~\rho_z\tau_z\sigma_x ~-~ \Delta_0 \rho_y\tau_0\sigma_y\ ,
		\end{align}
		
		As before, $m=m_0-2t$ is the effective Dirac mass and the $k_x^2$ term is neglected in $H_p$. Assuming a zero-energy trial solution for $H_0$ as, 
		\begin{equation}
			\psi_\beta(y) = C e^{\gamma y} \chi_\beta\ ,
		\end{equation}
		the equation $H_e \psi_\beta(y) = 0$ can be written as
		\begin{equation}
			\det\Bigg[ \left(m - \frac{t}{2} \gamma^2 \right)~ \rho_z\tau_z\sigma_0 ~-~ i \lambda_y \gamma ~ \rho_z\tau_y\sigma_0 \Bigg]=0\ .
		\end{equation}
		
		The corresponding solutions can be obtained as
		\begin{eqnarray}
			\gamma &= \frac{-\lambda_y-\sqrt{2mt+\lambda_y^2}}{t}\  ,
			\label{S24}
		\end{eqnarray}
		\begin{eqnarray}
			\gamma &= \frac{\lambda_y-\sqrt{2mt+\lambda_y^2}}{t} \  ,
			\label{S25}
		\end{eqnarray}
		\begin{eqnarray}
			\gamma &= \frac{-\lambda_y+\sqrt{2mt+\lambda_y^2}}{t} \  ,
			\label{S26}
		\end{eqnarray}
		\begin{eqnarray}
			\gamma &= \frac{\lambda_y+\sqrt{2mt+\lambda_y^2}}{t} \  .
			\label{S27}
		\end{eqnarray}
		
		Subject to the boundary conditions: $\psi_\beta(0)=0 ~;~\psi_\beta(\infty)=0$, only  Eq.~(\ref{S24}) and Eq.~(\ref{S26}) remain valid solutions. Substituting these $\gamma$ expressions back in the equation 
		$H_e \psi_\beta(x) = 0$, the normalized eigenspinors are obtained as
		\begin{equation}
			\chi_1=\frac{1}{\sqrt{2}}
			\begin{pmatrix}
				0\\0\\0\\0\\0\\1\\0\\1
			\end{pmatrix},
			\hspace{0.5cm}
			\chi_2=\frac{1}{\sqrt{2}}
			\begin{pmatrix}
				0\\0\\0\\0\\1\\0\\1\\0
			\end{pmatrix},
			\hspace{0.5cm}
			\chi_3=\frac{1}{\sqrt{2}}
			\begin{pmatrix}
				0\\1\\0\\1\\0\\0\\0\\0
			\end{pmatrix},
			\hspace{0.5cm}
			\chi_4=\frac{1}{\sqrt{2}}
			\begin{pmatrix}
				1\\0\\1\\0\\0\\0\\0\\0
			\end{pmatrix}.\ .
		\end{equation}
		
		Considering $m<0$ acoording to the Fu-Kane criterion~\cite{fu2007topological}, \ie $m_0 < 2t$ then yields,
		\begin{align}
			\gamma &= \kappa_1 \pm i \kappa_2\  , \\
			\kappa_1&=\frac{\lambda_y}{t} ~~;~~
			\kappa_2 =\sqrt{\frac{2|m|}{t}-\kappa_1^2}\ .  \nonumber  	
		\end{align}	
		
		Thus, the zero-energy trial solution is
		\begin{equation}
			\psi_\beta(y) = C e^{-\kappa_1 y} \sin(\kappa_2 y) ~ \chi_\beta  	\hspace{1.5cm} (\beta=1,2,3,4)\ ,
		\end{equation}
		here the normalization constant $C = \frac{2}{\kappa_2} \sqrt{\kappa_1 (\kappa_1^2 + \kappa_2^2)}$. 
		The perturbation $H_p$ can then be expressed using ${\psi_\beta(y)}$ as basis states
		\begin{equation}
			(H_{edge}^{II})_{\beta_1,\beta_2} =  \int_{0}^{\infty} \langle 	\psi_{\beta_1}(y) | H_p | \psi_{\beta_2}(y)\rangle  ~ dy 
			\hspace{1cm} (\beta_1,\beta_2 = 1,2,3,4)\ ,
		\end{equation}
		Therefore, the Hamiltonian for edge-II comes out to be
		\begin{equation}
			H_{\rm edge}^{\rm II} = -(\lambda_x+\alpha_x) k_x ~\tau_0\sigma_z - \Delta_0 ~\tau_y\sigma_y\ .
		\end{equation}
		
		For the edge-IV, identical procedure can be followed with the substitution $k_x \rightarrow -(-i\partial_x)$, and the corresponding edge Hamiltonian is then $H_{\rm edge}^{\rm IV} = H_{\rm edge}^{\rm II}$. This is expected due to the underlying $C_2$ symmetry being preserved in the heterostructure as mentioned before. 
		\subsection{Corner mode solutions}\label{Sec:II.C}
		As derived before, the Hamiltonian for edge-II is 
		\begin{equation*}
			H_{\rm edge}^{\rm II} = -(\lambda_x+\alpha_x) k_x ~\tau_0\sigma_z - \Delta_0 ~\tau_y\sigma_y\ ,
		\end{equation*}
		Employing OBC along $x$-direction leads to the substitution $k_x \rightarrow -i\partial_x$ and considering a trial solution at zero-energy $\psi_c \propto e^{-\zeta x} \phi_c$ , the secular equation reads
		\begin{equation}
			\det\left[-i \zeta (\lambda_x+\alpha_x) \tau_0\sigma_z - \Delta_0 \tau_y\sigma_y \right] = 0\ ,
		\end{equation}
		Then the values obtained are $\zeta = \big(\frac{-\Delta_0}{\lambda_x+\alpha_x} , \frac{\Delta_0}{\lambda_y+\alpha_x} \big)$. The boundary condition $\psi_c(0)=0$ requires the $\zeta>0$ to be the acceptable solution. Substituting back this $\zeta$ value into the equation $H_{\rm edge}^{\rm II} \psi_c=0$ 
		with $\phi_c=(x_1 , x_2 , x_3 , x_4)^T$ gives 
		\begin{equation}
			x_3=i x_2  ~~~;~~~ x_4=i x_1 
			\label{S34}
		\end{equation}
		
		On the other hand, the Hamiltonian for edge-I is 
		\begin{equation*}
			H_{\rm edge}^{\rm I} = (\lambda_y-\alpha_y) k_y ~\tau_0\sigma_z + J_{sd} ~\tau_z \sigma_z - \Delta_0 ~\tau_y\sigma_y\ ,
		\end{equation*}
		Imposing OBC along $y$-direction leads to the substitution $k_y \rightarrow -i\partial_y$ and considering a trial solution at zero-energy $\psi_c \propto e^{-\zeta y} \phi_c$ , the secular equation reads
		\begin{equation}
			\det\left[i \zeta (\lambda_y-\alpha_y) \tau_0\sigma_z + J_{sd} \tau_z\sigma_z - \Delta_0 \tau_y\sigma_y \right] = 0\ ,
			\label{S35}
		\end{equation}
		The corresponding values obtained are
		\begin{align*}
			\zeta_1&= \frac{-J_{sd}-\Delta_0}{\lambda_y-\alpha_y} ~~;~~ \zeta_2 = \frac{J_{sd}-\Delta_0}{\lambda_y-\alpha_y} ~~;~~
			\zeta_3 = \frac{-J_{sd}+\Delta_0}{\lambda_y-\alpha_y} ~~;~~ \zeta_4 = \frac{J_{sd}+\Delta_0}{\lambda_y-\alpha_y} 
		\end{align*}
		
		When $J_{sd}<\Delta_0$, the boundary condition $\psi_c(0)=0$ dictates $\zeta_3$ and $\zeta_4$ to be the acceptable solutions. Solving $H_{\rm edge}^{\rm I} \psi_c=0$ with $\phi_c=(y_1 , y_2 , y_3 , y_4)^T$ and $\zeta_3$, we obtain
		\begin{equation}
			y_2=i y_1  ~~~;~~~ y_3= y_1 ~~~;~~~ y_4=-iy_1\ ,
			\label{S36}
		\end{equation}
		Similarly, with $\zeta_4$,	
		\begin{equation}
			y_2=-i y_1  ~~~;~~~ y_3= -y_1 ~~~;~~~ y_4=-iy_1\ ,
			\label{S37}
		\end{equation}
		Comparing Eq.~(\ref{S34}), Eq.~(\ref{S36}) and Eq.~(\ref{S37}), same solutions can never be obtained for any values of $(x_{1,2,3,4},y_{1,2,3,4})$. Thus, no MCMs are possible for $J_{sd}<\Delta_0$.

		In case when $J_{sd}>\Delta_0$, the boundary condition $\psi_c(0)=0$ guarantees $\zeta_2$ and $\zeta_4$ to be the acceptable solutions. Solving $H_{\rm edge}^{\rm I} \psi_c=0$ with $\phi_c=(y_1 , y_2 , y_3 , y_4)^T$ and $\zeta_2$, we obtain
		\begin{equation}
			y_2=-i y_1  ~~~;~~~ y_3= y_1 ~~~;~~~ y_4=iy_1\ ,
			\label{S38}
		\end{equation}
		Thus comparing Eq.~(\ref{S34}) and Eq.~(\ref{S38}), a common solution $\phi_c=(1,-i,1,i)^T$ is obtained.

		Hence, one zero-energy Majorana mode per corner can be obtained only when $J_{sd}>\Delta_0$. This conclusion is also supported by the analysis of quadrupolar winding number $\mathcal{N}_{xy}$, where $\mathcal{N}_{xy}=1$ in the region when $J_{sd}>\Delta_0$ (see Fig.~4(b) of the main text). 
		
		\subsection{Jackiw-Rebbi condition}\label{Sec:II.D}
		The previous discussions considers the QSHI in the exact part $H_0$ and rest of the terms from $H_{\rm eff}$ remain in the perturbation $H_p$, thereby leading 
		to the condition $J>\Delta_0$ for appearance of the MCMs. To obtain a deeper insight, we incorporate the spin-split hopping terms $\alpha_{x,y}$ from the $p$WM into the $H_0$ and proceed in the similar manner as discussed before. 
		This leads to the modified Hamiltonians for edge-I and II (same for edge-III and IV) as
		\begin{align}
			H_{\rm edge,\alpha}^{\rm I} &= J \eta_x \sqrt{1-\eta_x} ~\tau_0\sigma_y + k_y (\lambda_y \sqrt{1-\eta_x^2}-\alpha_y) ~\tau_0\sigma_z + J(1-\eta_x^2) ~\tau_z\sigma_x - \Delta_0 \tau_y\sigma_y\ , \\
			H_{\rm edge,\alpha}^{\rm II} &= (-J\eta_y\sqrt{1-\eta_y^2})~\tau_0\sigma_y - k_x (\alpha_x + \lambda_x\sqrt{1-\eta_y^2})~\tau_0\sigma_z + J \eta_y^2 ~\tau_z\sigma_x - \Delta_0 ~\tau_y\sigma_y\ .
		\end{align} 
		Here, $\eta_{x,y}=\frac{\alpha_{x,y}}{\lambda_{x,y}}$. Evaluating the eigenvalues at $k_{x,y}=0$, the mass gaps for the edges-I and II are 
		\begin{align}
			g_1 &= (J\sqrt{1-\eta_x^2}-\Delta_0)\ , \\
			g_2 &= (J\eta_y-\Delta_0)\ .
		\end{align}
		Hence the Jackiw-Rebbi condition~\cite{jackiw1976solitons,ghosh2021floquet,shen2012topological} appears as $g_1 \cdot g_2 < 0$ \ie
		\begin{equation}
			\boxed{(J\sqrt{1-\eta_x^2}-\Delta_0)(J\eta_y-\Delta_0) < 0}\ .
		\end{equation}
		This also indicates the restriction $\lambda_x > \alpha_x$ as $\eta_{x}<1$. Assuming $\lambda_x=\lambda_y$, and $\alpha_x=\alpha_y$, we obtain a region in the parameter space of $(J_{sd},\Delta_0)$ in which the system hosts SOTSC phase anchoring localized MCMs (refer to the main text for discussions).
		
		\section{DERIVATION OF EFFECTIVE BULK PAIRINGS}\label{Sec:III}
		Here we derive the effective bulk pairings of our setup as discussed in the main text. This requires a duality transformation of the effective Hamiltonian $H_{\rm eff}$, given in Eq.~(\ref{Heff}), to a dual Hamiltonian $H_D$ through a dual matrix $U_D$~\cite{sato2009non}.
		Following the procedure described in Ref~\cite{chatterjee2024second}, we consider the duality matrix $U_D$ after a unitary transformation $P$, owing to the different bases:
		\begin{align}
			&\left(c_{\boldsymbol{k}A\uparrow} ~,~ c_{\boldsymbol{k}A\downarrow} ~,~ c_{\boldsymbol{k}B\uparrow} ~,~ c_{\boldsymbol{k}B\downarrow} ~, -c_{\boldsymbol{-k}A\downarrow}^\dagger ~,~ c_{\boldsymbol{-k}A\uparrow}^\dagger ~, -c_{\boldsymbol{-k}B\downarrow}^\dagger ~,~ c_{\boldsymbol{-k}B\uparrow}^\dagger \right)^T \nonumber \\
			=~& P  \left(c_{\boldsymbol{k}A\uparrow} ~,~ c_{\boldsymbol{k}A\downarrow} ~,~ c_{\boldsymbol{k}B\uparrow} ~,~ c_{\boldsymbol{k}B\downarrow} ~,~ c_{\boldsymbol{-k}A\uparrow}^\dagger ~,~ c_{\boldsymbol{-k}A\downarrow}^\dagger ~,~ c_{\boldsymbol{-k}B\uparrow}^\dagger ~,~ c_{\boldsymbol{-k}B\downarrow}^\dagger \right)^T\ .
		\end{align}
		Here, $P=\begin{pmatrix}
			\tau_0\sigma_0 & 0 \\ 0 & -i\tau_0\sigma_y 
		\end{pmatrix}$.
		Then the dual matrix in the basis $\Psi_{\boldsymbol{k}}$ (Eq.~(\ref{Psib})) is given by
		\begin{align}
			U'_D &= P^\dagger U_D P \nonumber \\
			&= \frac{1}{\sqrt{2}} (\rho_0\tau_0\sigma_0 + i \rho_x\tau_0\sigma_y)\ ,
		\end{align}
		Here, $U_D = \frac{1}{\sqrt{2}}\begin{pmatrix}
			1 & -1 \\ 1 & 1
		\end{pmatrix} \tau_0\sigma_0 $. Note that, the duality matrix $U'_D$ obtained here is identical to the one as mentioned in Ref~\cite{sato2009non}. The dual Hamiltonian is then obtained by
		\begin{gather}
			H_D = (U'_D)^\dagger H_{\rm eff} U_D \nonumber \\
			~~~~= \begin{pmatrix}
				\epsilon_D & \Delta_D \\ \Delta_D^\dagger & -\epsilon_D
			\end{pmatrix}\ ,
		\end{gather}
		where, $\epsilon_D = \Delta_0 ~\tau_0\sigma_0 + J_{sd} ~\tau_x\sigma_x$ and $\Delta_D = \Delta_s + \Delta_{p_x+p_y} + \Delta_{p_x+ip_y}$.
		\begin{align}
			\Delta_s &= -i(m_0 - 2t) \tau_z\sigma_y\ , \nonumber \\
			\Delta_{p_x+p_y} &= -(\alpha_x k_x + \alpha_y k_y) ~\tau_0\sigma_x\ , \nonumber \\
			\Delta_{p_x+ip_y} &= -(\lambda_x k_x ~\tau_x\sigma_x + i \lambda_y k_y ~\tau_y\sigma_y)\ .
			\label{DP}
		\end{align}
		
		The minus signs in Eq.~(\ref{DP}) appear due to the choice of basis. Here, $(p_x+p_y)$ pairing is known to exhibit Majorana flat edges modes (MFEMs)~\cite{nakosai2013two,chatterjee2024topological,wang2017topological,zhang2019majorana} in contrast to $(p_x+ip_y)$ pairing. The latter exhibits dispersive Majorana edge modes~\cite {nakosai2013two,hu2019chiral}. The nature of edge modes in our system can be modulated by the relative magnitudes of $\lambda_{x,y}$ and $\alpha_{x,y}$ at low-energy and near $\boldsymbol{k}\approx 0$. In the limit $\alpha_{x,y} \approx 0$, the system behaves as a QSHI and first order topology is observed on all four edges due to $C_4$ symmetry. The edges are trivially gapped in presence of $s$-wave pairing. Similarly, for $\lambda_{x,y} \approx 0$, effective bulk $(p_x+p_y)$ pairing leads to FOTSC hosting MFEMs at opposite edges~\cite{pal2026emergent} as they preserve $C_2$ symmetry. Therefore, both $\lambda_{x,y}$ and $\alpha_{x,y}$ are required to be non-zero so as to obtain MCMs (SOTSC) as evident from our numerical analysis 
		and low energy edge theory.
		
		\section{CHIRAL SYMMETRY AND TOPOLOGICAL INVARIANT}\label{Sec:IV}
		The Hamiltonian $H_{\rm BdG}$ possesses chiral symmetry $\mathcal{S}=\rho_0\tau_x\sigma_x$. This enables us to unitarily transform $H_{\rm BdG}$ into an anti-diagonal form $\mathcal{H}=\big( \begin{smallmatrix} 0 & q(\boldsymbol{k})\\ q(\boldsymbol{k})^\dagger & 0 \end{smallmatrix}\big )$, where \\
		\begin{center}
			{\scriptsize $q(\boldsymbol{k}) = \begin{bmatrix}
					\begin{aligned}[t]
						m_0  &\\ - t (\cos k_x+t \cos k_y) &\\ -\alpha_x \sin k_x -\alpha_y \sin k_y
					\end{aligned} & \begin{aligned}[t]
						J_{sd}-\lambda \sin k_x &\\ -i \lambda \sin k_y 
					\end{aligned}& 0 & \Delta \\ \\ \\
					
					\begin{aligned}
						J_{sd}+\lambda \sin k_x &\\ -i \lambda \sin k_y 
					\end{aligned} & \begin{aligned}[t]
						m_0  &\\ - t (\cos k_x+ \cos k_y) &\\ +\alpha_x \sin k_x +\alpha_y \sin k_y
					\end{aligned} & -\Delta & 0 \\ \\ \\
					
					0 & -\Delta & \begin{aligned}[t]
						-m_0  &\\ + t(\cos k_x+ \cos k_y) &\\ -\alpha_x \sin k_x -\alpha_y \sin k_y
					\end{aligned} & \begin{aligned}[t]
						-J_{sd}-\lambda \sin k_x &\\ +i \lambda \sin k_y 
					\end{aligned} \\ \\ \\
					
					\Delta & 0 & \begin{aligned}[t]
						-J_{sd}+\lambda \sin k_x &\\ +i \lambda \sin k_y 
					\end{aligned} & \begin{aligned}[t]
						-m_0  &\\ + t (\cos k_x+ \cos k_y) &\\ +\alpha_x \sin k_x +\alpha_y \sin k_y 
					\end{aligned} \\	
				\end{bmatrix}$}\ .
		\end{center}
		\vspace{0.4cm}
		
		This leads us to define one-dimensional $\mathbb{Z}$ topological invariant- the winding number $\mathcal{N}_y (k_x)$ as mentioned in the main text. However, 
		this can only characterize the first-order topological phase and therefore a different topological invariant is required for the second-order phase. To this end, a quadrupolar winding number $\mathcal{N}_{xy}$ can be defined in the real-space~\cite{benalcazar2022chiral,pal2025multi,subhadarshini2025engineering} as discussed in the main text. There, $U_S$ is the eigenvector matrix for the chiral operator $\mathcal{S}$. The transformed Hamiltonian $\mathcal{H}$ is in the chiral basis, corresponding to the chiral eigenvalues +1 and -1 for the A and B sublattice sectors, respectively. Using the real-space version of the off-diagonal matrix $q$, a singular value decomposition (SVD) is performed as $q = U_A \Sigma U_B^\dagger$, where $\Sigma$ is a diagonal matrix in singular values. Also, $U_{A},U_{B}$ are the singular vector matrices that together form an orthonormal basis for the chiral operator. The invariant $\mathcal{N}_{xy}$ indicates the second-order bulk-topology and reflects the number of localized MCMs at each corner of the 2D domain.


	\end{onecolumngrid}

\end{document}